\documentclass[aps,twocolumn]{revtex4-1}
\usepackage{amsmath,amssymb}
\usepackage{slashed}
\usepackage{graphicx}
\usepackage{bm}
\usepackage[T1]{fontenc}
\usepackage[utf8]{inputenc}
\usepackage{gauss} 
\usepackage[top=1.0in,bottom=1.0in,left=1.0in,right=1.0in]{geometry}
\usepackage[colorlinks,linkcolor=blue,citecolor=blue]{hyperref}
\usepackage{subfig}
\newcommand{\be}{\begin{equation}}
\newcommand{\ee}{\end{equation}}
\newcommand{\bea}{\begin{eqnarray}}
\newcommand{\eea}{\end{eqnarray}}
\newcommand{\ba}{\begin{eqnarray}}
\newcommand{\ea}{\end{eqnarray}}

\begin{document}

\title{Mesonic ``screening masses" in high temperature QCD}

\author{Edward Shuryak}
\email{edward.shuryak@stonybrook.edu}
\affiliation{Center for Nuclear Theory, Department of Physics and Astronomy, Stony Brook University, Stony Brook, New York 11794--3800, USA}

\begin{abstract}
Lattice studies of QCD at high temperature has investigated the so called ``screening masses",
corresponding to quark-antiquark states propagating in $spatial$ (rather than time) direction.
Thirty years ago we pointed out that those should correspond to states of 2+1 dimensional
quarkonia, and in this work we show that this correspondence provide quantitative
description of the lattice data.
 \end{abstract}
\maketitle

\section{Introduction}
 Hadronic correlation functions at nonzero temperatures, measured  in time
 or space directions, are different. The logarithmic derivatives of those give us
 {\em hadronic masses} and {\em screening masses}, respectively.  
 
 The pioneering lattice studies \cite{DeTar:1987ar} found that, at certain temperature
 above the deconifnement critical temperature $T>T_c$, the extracted
rho-meson {\em screening mass} was close to the original $mass$ at zero temperature.
While still realizing  that their ``screening masses" are not really masses,  DeTar and Kogut still ended the paper by noticed  that "...their appearance in the screening spectrum deals a serious blow to the naive deconfinement picture...". As we will see below, confusions of this kind periodically
 reappear to this day.
 
Thirty years ago it was pointed out in our paper \cite{Koch:1992nx} 
that these observations have nothing to do with deconfinement.
 It is rather trivial to see that mesonic screening masses should have a discrete spectrum   at
 $all$ temperatures, ranging to $T\rightarrow \infty$, since in this limit
the theory is just becomes the 3-dimensional QCD, known to be a confining theory.

%General theory discussion of mesonic screening masses at high $T$ started by 
It was pointed out that QCD defined on Euclidean $R^3C^1$ manifold has two identical
intepretations:\\
In the {\bf formulation  A} the  compact coordinate is the
 Euclidean time $x_4$, therefore the circumference of $C^1$ is identified with inverse
 temperature $T$. Correlators in the direction $x_3$ give screening masses.\\
In the  {\bf formulation B}  the  compact coordinate is called $x_3$, while the time $x_4$ is set to be one
of the non-compact directions. So the theory is at zero temperature, but interpolating between
the {\em four-dimensional} QCD at $T\rightarrow 0$ and the  {\em three-dimensional} QCD at $T\rightarrow \infty$. 

Antiperiodic boundary conditions for quarks on the circle $C^1$ result in 
corresponding momentum quantized into odd  ``Matsubara frequencies" $\pm\pi T,\pm 3\pi T, ...$
growing with $T$. Therefore, at high enough $T$ in the  formulation {\bf B}, the problem reduces to a {\em nonrelativistic quarkonium} problem. Note that since $\pi T$ is then
interpreted as the heavy quark mass. (It would match charm quark mass ($m_c\approx 1.5\, GeV$)
at $T\sim 3T_c\sim 0.5\, GeV$. ) Therefore, mesons in the high $T$ domain  may be approximated by the states of the {\em 2+1 dimensional quarkonium}. Following these ideas, one can
quantitatively reproduce recent lattice data on the screening masses.

In a number of recent works, from
\cite{Glozman:2019fku} to \cite{Glozman:2022lda}, lattice data on screening masses
were erroneously related with ``absence of deconfinement". It is further claimed 
that there is establishment of a
novel {\em chiral spin symmetry} for $T_c<T<3T_c$ corresponding to suppression of spin-spin forces.
Here is not a place for criticism, so let me just say these papers have no quantitative results
neither on screening masses, nor vector-scalar splittings induced by spin-spin forces.

Let us return to quarkonia, recalling crucial development in hadronic spectroscopy in 1970's. 
Following
the discovery of multiple quarkonia states of $\bar c c $ and $\bar b b$ families, it was shown that their spectra can be accurately reproduced with potentials
having only three basic elements:\\
(i) pertubative Coulomb term corresponding to one gluon echange;\\
(ii) nonperturbative confining central potential linear in $r$;\\
(iii) $O(1/M^2)$ suppressed spin-dependent potentials, of spin-spin, spin-orbit and tensor kinds.\\ 

The aims of this paper are very similar to that development. Given now available lattice-based screening mass spectrum, one may figure out the properties of 
the underlying effective potential. With the information
at hand, we can basically quantify  parameters of the linear and quadratic potential, as well  as the magnitude
of quasi-local spin dependent force splitting vector/axial and  scalar/pseudoscalar channels.
These resulting potentials can be used, in many-body context, to calculate thermal
and kinetic properties of QGP at high $T$.

\begin{figure}[h]
\begin{center}
\includegraphics[width=7.5cm]{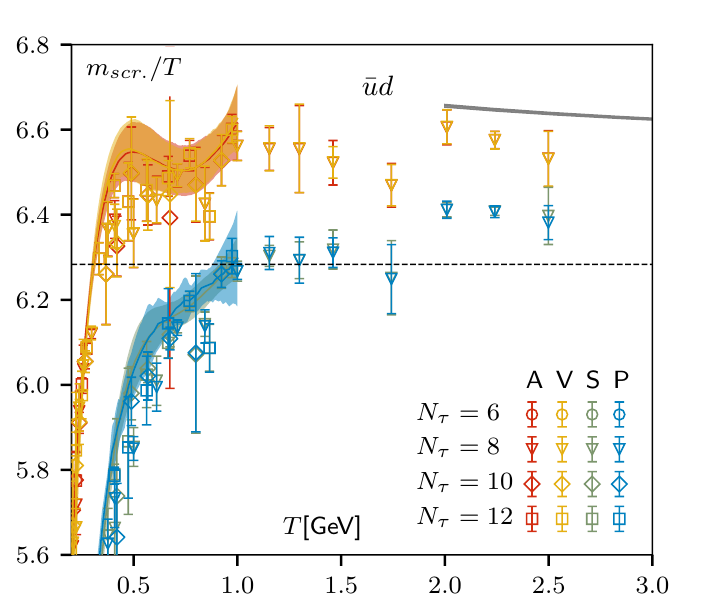}
\includegraphics[width=8.5cm]{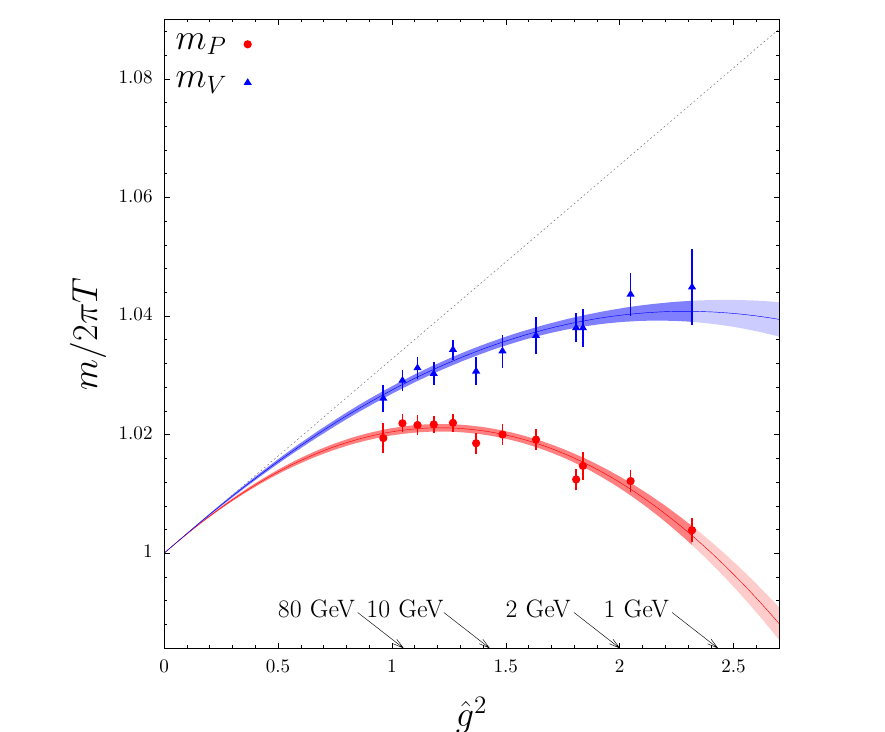}
\caption{In the upper plot mesonic screening masses are given  normalized to the temperature $m_{scr}/T$
versus $T\, (GeV)$. In the lower plot the nomalization is $m_{scr}/2 \pi T$, plotted versus the effective coupling $\hat g$.  Blue points and curve are for vector/axial channel, lower red ones f
are or scalar/pseudoscalar channels. }
\label{fig_screening_mesons_hiT}
\end{center}
\end{figure}

%\section{Recent input from the lattice }

Recently screening masses were calculated in wider range of temperatures/couplings,
by \cite{Bazavov:2019www}, and then by
 \cite{Laudicina:2022ghk}. Their results are
  reproduced as upper and lower plots of  Fig.1. 
Note that the latter results are for quite wide temperature range,
 from $T=1 - 160\, GeV$. The corresponding
 coupling range is $\hat g^2 =2.7-1.0$, so naively one may doubt
 validity of the perturbative analysis. Yet, as we will show in the next
 section, it actually works quite well.

%The main features of the results shown in Fig.\ref{fig_screening_mesons_hiT}
%are as follows:
%
%(i) in the high temperature domain, at  $T>1\,GeV$ the screening masses
%(normalized to $T$) 
%of axial/vector mesons are approximately stabilized to a value  

\section{Theory of the  2+1-dimensional ``quarkonium"}
The 3-dimensional
effective theory starts with observation that to one-loop order the gluon
gets  the so called {\em electric screening mass} found in \cite{Shuryak:1977ut}
\be m_E^2=({N_c\over 3}+{N_f \over 6}) g^2T^2
\ee
For brevity, we will below consider only the case $N_c =N_f=3$.

A dimensionful 3-D electric coupling is  $g_E^2=g^2T$, and one can imagine
effective Lagrangian for quarks containing terms of different order in this coupling.

Quantitative studies of the high-T screening masses have been pioneered 
in \cite{Laine:2003bd}, which we follow
in selecting units and effective potential.  If
the radial coordinate is normalized to inverse electric mass  \be \hat r= r M_E \ee 
one get effective Schreodinger equation 
 (eqn  (6.15) of \cite{Laine:2003bd}) which has the form 
\ba \label{eqn_Schr}
 -\psi''-{1\over x} \psi'+\rho\big(log(x/2)+\gamma_E-K_0(x) \big)\psi \nonumber \\ 
 =\hat E_0 \psi %
 \ea
Note that the coefficient in front of the potential 
 %\psi(x) $$ \be +{(\pi T)(C_M g^4 T^2) \over  M_E^3} x \psi(x)=\epsilon \psi(x) \ee
\be \rho={p_0 g_E^2 C_F \over 2\pi m_E^2}={3(N_c^2-1) \over 2N_c (2N_c+N_f)}\ee
is $universal$, independent on both $T$ and the coupling $\hat g^2$. For the case we need, 
$N_c =N_f=3$, $\rho=4/9$. Here $C_F=4/3$ is the standard color factor for quark-antiquark interactions.

The "quarkonium" mass is close to twice the Matsubara  frequency
 $\pi T$, but in this eqn it is without two because, as usual, it is the reduced mass for relative motion. The universal eigenvalues of this operator are related to the quarkonium mass via
 \be \label{eqn_Mfull}
 M_{full}=2\pi T+g^2 T {C_F \over 2\pi} ({1\over 2}+ \hat E_0) \ee
 so the actual deviation from asymptotic value is linear in coupling $g^2(T)$, for which
 the usual two-loop expression is used 
 \be {1 \over g^2(T)}={9 \over 8\pi^2}log({2\pi T \over \Lambda})+{4 \over 9\pi^2}log\big( 2 log({2\pi T \over \Lambda})\big)
 \ee
 The value $\Lambda \approx 0.5\, GeV$ leads to coupling values as used in \cite{Laudicina:2022ghk}.
 
 The equation $(\ref{eqn_Schr})$ does not have analytic solution, but can easily be solved numerically
 (we do not know why that was not done in \cite{Laine:2003bd}). 
 
\begin{figure}[h]
\begin{center}
\includegraphics[width=8cm]{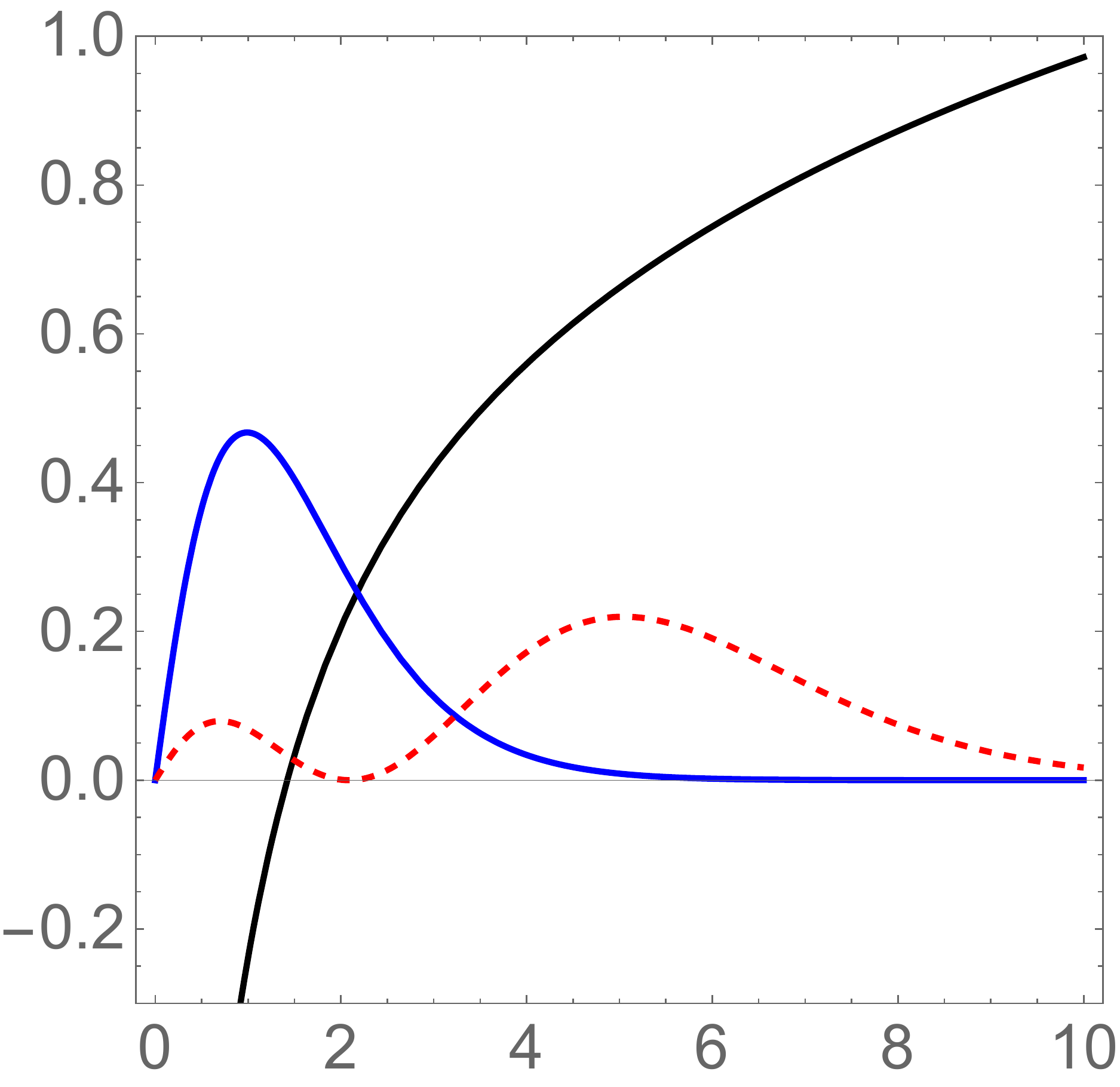}
\caption{The effective potential is shown by the black solid curve, shown together with
probability distribution for the ground state
$2\pi r | \psi_0(\hat r)|^2$ (blue solid) and for the first excited state $2\pi r |\psi_1(\hat r)|^2$ (red dashed).
}
\label{fig_pot_two_levels}
\end{center}
\end{figure}

 In Fig.\ref{fig_pot_two_levels}
 we show the 2d Coulomb potential, together with the probabilities to find system in the ground and the first excited states. It shows that
 those spatial distributions are quite different. While for the ground state the maximum is
 close to zero of the potential, this is not so for other states. As a result, the energies of these states are quite different, 
 \be \hat E_{0,1,2}=0.208618, 0.836333, 1.07963 ...
 \ee
 with that of the ground state much smaller than one because of partial cancellation 
 of large and 
 small $\hat r$ regions, where the potential is of opposite sign.  According to (\ref{eqn_Mfull})
 \be  {M_{full}\over 2\pi T}=1+g^2 {C_F \over 4\pi^2} (0.5+0.208618)=1+g^2 0.0239327
 \ee
 which does not agree with numerical coefficient suggested in \cite{Laine:2003bd}, 
$1+g^2 0.032739961$ (based on some
 estimates of the wave function near the origin). Therefore, we find that the dashed 
 line in the lower Fig.\ref{fig_screening_mesons_hiT} should have different, significantly smaller, slope. As we will see soon, this
 change actually brings it much closer to the lattice data.
 
%%%%%%%%%%%%%%
\begin{table}[htp]
\caption{The average values of the reduced $\hat r$, $\hat r^2$ and 2-d delta function
for the first two levels of 2+1 dimensional quarkonia.}
\begin{center}
\begin{tabular}{|c|c|c|c|}
\hline
& $\langle n | \hat r | n  \rangle$ & $\langle n | \hat r^2 | n \rangle$ & $\langle n| \delta^2( \hat r) | n \rangle$ \\ \hline 
n=0 & 1.58652 & 3.49549 & 0.147316 \\
n=1 & 5.32307 &  33.0913 & 0.0320479 \\
\hline
\end{tabular}
\end{center}
\label{tab_higher}
\end{table}%

Using the obtained wave functions, one can also perturbatively evaluate
effects due to higher order in coupling in the Schreodinger equation
\be \hat V_{higher}=C_4 g^2 \hat r+ C_6  g^4 \hat r^2+... \ee  
(Note that we use as the index for coefficients not the power of coupling in this equation, but
the power of $g$ in the full mass.) 

In the next orders there should appear spin-dependent forces which would split
vector and scalar quarkonia.
One may also introduce (in the simplest local form) the 
spin-dependent potential, producing shift in the mass of scalar/pseudoscalar states
\be  \hat V_{SS}=C_{spin} g^2\big({1-\vec\sigma_1 \vec\sigma_2\over 2}\big) \delta^2(\hat r)
\ee
The Table \ref{tab_higher} give the corresponding values of mean $\hat r, \hat r^2,\delta^2(\hat r)$
for the first two states. Using those and the data from \cite{Laudicina:2022ghk},one can fit
the coefficients of the higher order  terms.
For the spin term we follow a fit from \cite{Laudicina:2022ghk} 
\be  {m_V-m_S \over 2\pi T}= 0.00704*g^4
\ee
from which and Table 1 the  $C_{spin}$ value follows.

\begin{figure}[t]
\begin{center}
\includegraphics[width=8cm]{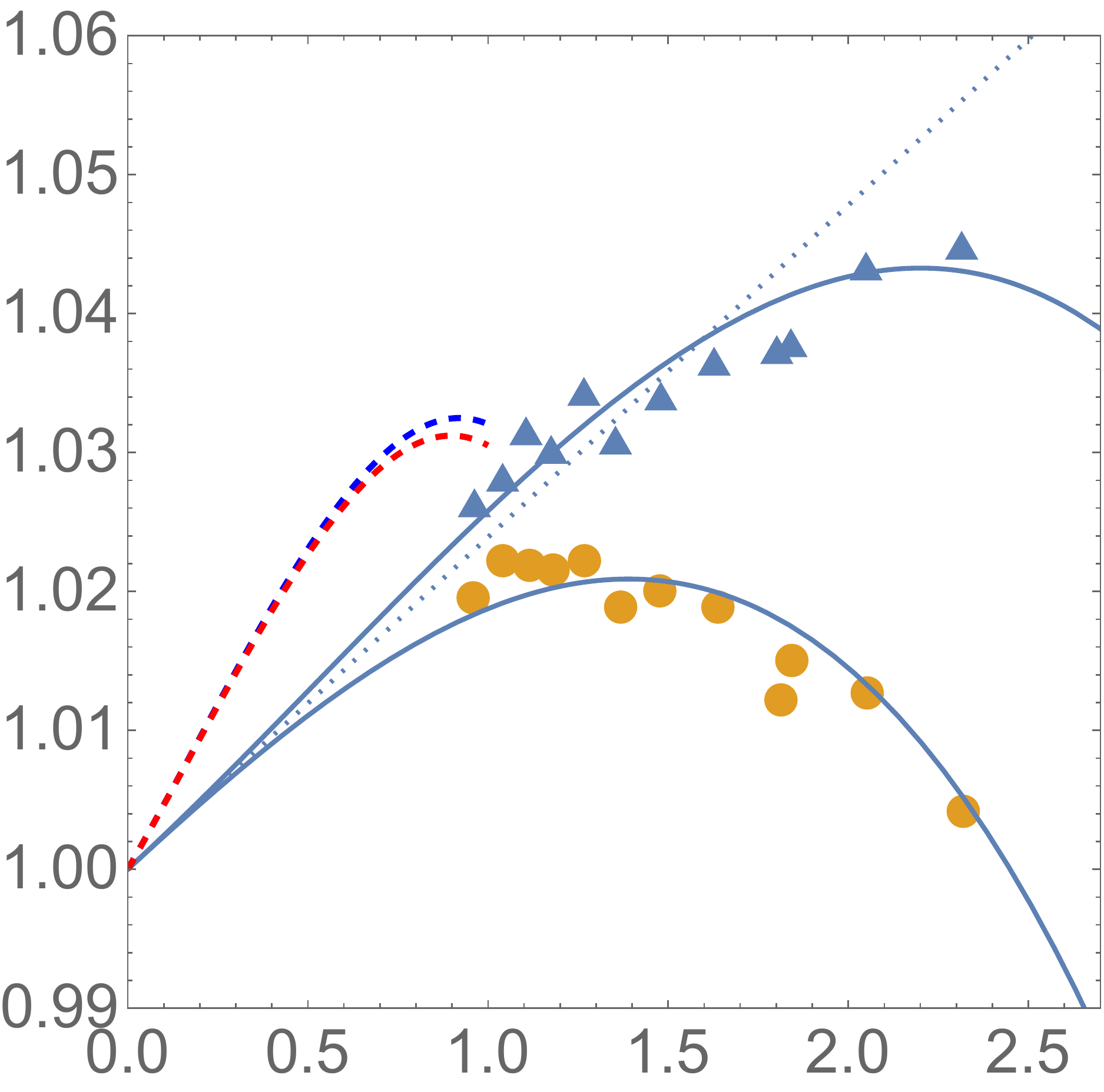}
\caption{Normalized mesonic screening masses  $m_{scr}/2 \pi T$ 
 versus squared effective coupling $\hat g^2$. The dotted straight line is the corrected
 $O(\hat g^2)$ asymptotic term. Blue lines are for vector (upper) and scalar (lower)
 ground states, with parameters given in the text. 
  Blue triangles  and  red discs show vector and scalar masses, respectively:
  they are the same lattice data as in Fig.1. Finally, blue and red dashed lines in the left side
  of the plot show our predictions for the first excited vector and scalar states. }
\label{fig_new_masses}
\end{center}
\end{figure}

Revised version of the plot, compared to the same lattice data \cite{Laudicina:2022ghk}, is shown in Fig.\ref{fig_new_masses}
for the ground and first excited states in both V and S channels. 
The predictions for excited states are obtained perturbatively, based on the mean values 
given in the Table I. 
The extracted coefficients of the higher order terms in the effective action used in this plot are 
\be C_4 =.095, C_6 = -0.027
\ee
Note that in our parameterization the string tension (coefficient $C_4$ of $\hat r$)
is positive, unlike the one in the parameterization used in \cite{Laudicina:2022ghk}.

%\begin{widetext}
%\end{widetext}

\section{Comments on  the $[T_c,3T_c]$ strip}
According to Glozman et al \cite{Glozman:2019fku,Glozman:2022lda}  this temperature range
is ``stringy plasma", while at higher $T>3T_c\sim 0.5\, GeV$ the matter is 
in a ``true QGP" phase. The motivation for this
was that in the former case these authors were able to see that screening masses are different
from those for free quarks,  $M-2\pi T\neq 0$, while in the latter they could not see
this difference.

With simple theory arguments (as well
as better lattice data we discussed above) it is clear that these arguments are false.
The boundary proclaimed was simply dependent on the statistical accuracy
of the particular simulation used.
The shifts of screening masses remains nonzero $M-2\pi T\neq 0$ at any $T$.
The theoretical calculations of the previous section provide quantitative 
description of those in the ``quarkonium domain", roughly $T>1\, GeV$.

Yet it is true that inside this strip,  $[T_c,3T_c]$  one finds screening masses
qualitatively different from those at high $T$, in particular their deviation from free, $M-2\pi T$, 
gets
 negative at smaller $T$, see upper plot of Fig.1. Here we would like to relate this phenomenon to $qualitative$
 changes in the nature of the topological solitons, occurring in this range of  temperatures.
 
\begin{figure}[htbp]
\begin{center}
\includegraphics[width=8cm]{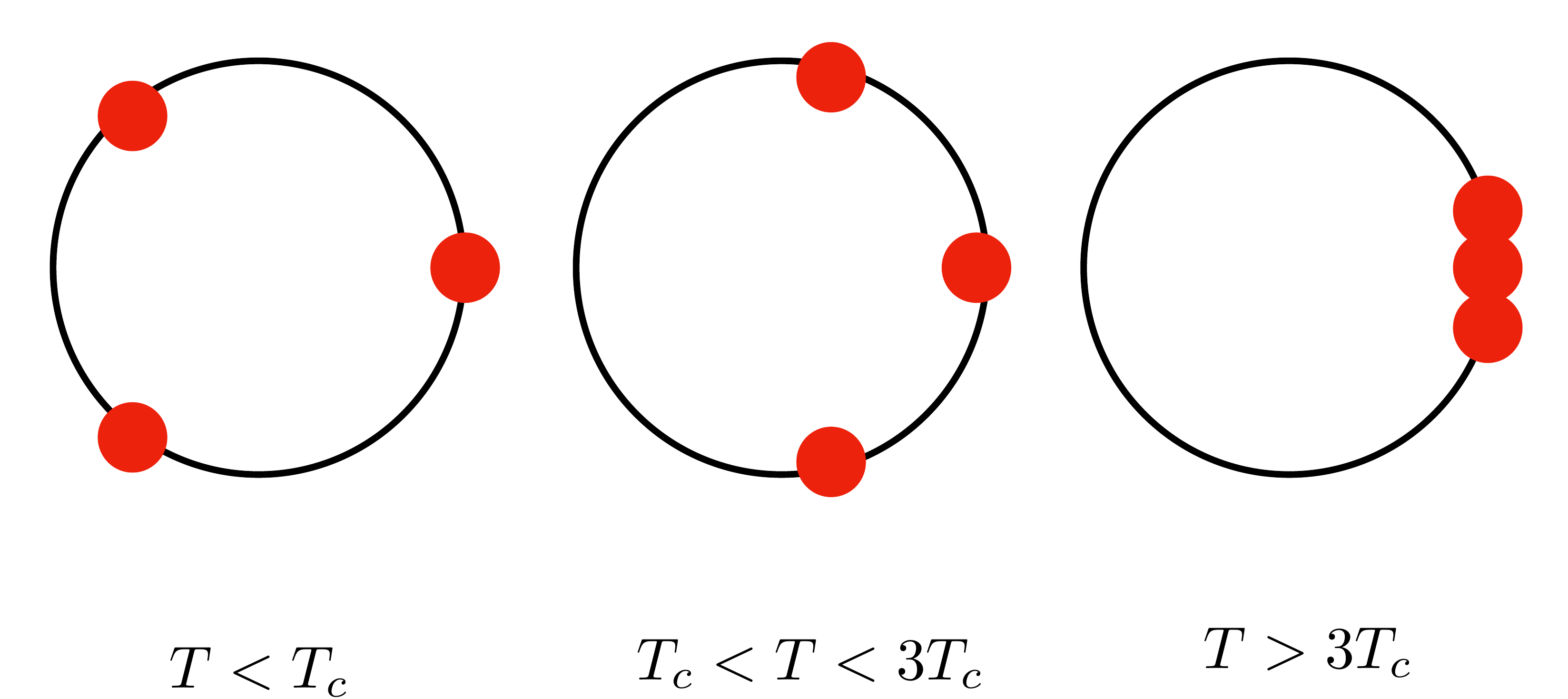}
\caption{Locations of the Polyakov line eigenvalues, below $T_c$, between $T_c$ and $3T_c$
and around $3T_c$ (and above)}
\label{fig_3circle}
\end{center}
\end{figure}

It is well known that the eigenvalues of the Polyakov line depend on temperature.
 In Fig.\ref{fig_3circle} we show their locations below the ``strip" in question, inside  and above
it. These three pictures correspond to mean values of the Polyakov line $\langle P\rangle =0$, between zero and one, and close to one at and above $3T_c$.  

What is also known (although not so widely) is that eigenvalues (three red dots)
divide the unit circle into three  sectors, and the actions of three types of {\em instanton-dyons}
are proportional to their lengths. One of them, called $L$ dyon, correspond to the left-side sector
between two complex points: it fraction of the circle changes from 1/3 at $T<T_c$ (left sketch), to 1 at   $T\approx 3T_c$ (right sketch). Two other dyons called $M_1,M_2$ take complementary
fractions, and together their actions satisfy the sum rule $S_L+S_{M1}+S_{M2}=8\pi^2/g^2$.

A relation between the topological solitons and quark-antiquark potential is based on the
fact that $L$ instanton-dyon possess fermionic zero mode, leading to 
nonperturbative quark-antiquark interactions described by
the so called 't Hooft
effective Lagrangian. (Dyons   $M_1,M_2$ do not have zero modes and are thus irrelevant.)
 The action $S_L$ changes dramatically in the  $[T_c,3T_c]$ strip, from about $\approx 4$
 on its lower end, to about $\approx 12$ at its high end. In leading order semiclassical
 approximation, their density thus should change be a  factor $exp(-\Delta S_L)=exp(-8)$.
 One-loop effects and interactions between dyons makes this drop more temperate.
 Unfortunately there are no quantitative data on that so far.
 Existing simulations \cite{DeMartini:2021xkg}  were done mostly in the chirally broken phase $T<T_c$, reaching only from $1.0$ to $1.15 T_c$ into the chirally restored phase. In this
 small temperature interval the density of $L$ dyons drops roughly by factor two.
 
 (Let us for clarity remind the reader, that at $T<T_c$ the $L$-dyon density is large enough to  break spontaneously chiral $SU(N_f)_c$ symmetry, forming quark condensate, making pions massless, etc. At  $T>T_c$ , in the setting with massless quarks and exact chiral symmetry, only ``neutral" clusters of total topological
 charge zero should be present, e.g. $\bar L L$ pairs. Such clusters generate ``squared" 't Hooft
 operators, or four-fermion operators of the NJL type.)
 
 So, while we do not provide  any quantitative calculation of the effect yet,
 we do suggest that rapid change in the screening masses in the discussed strip
(causing in particular
 a switch to negative $M-2\pi T$) is due to rapid change of the  $L$-dyon density
 in this temperature range.

\section{Summary}
Our results (curves) are summarized in Fig.\ref{fig_new_masses} and compared to lattice data. The main correction over the previous works
is that we did solved the leading order Schreodinger equation for 2+1 dimensional quarkonium, for the ground and few
excited states. With higher order effects added perturbatively, 
 it does describe well the available lattice data at $T>1\, GeV$. Note that $O(g^4)$ string tension
 is now found to be positive. To test this theory further one may try to perform lattice
 measurements of the first excited states, for which there are firm predictions at high $T$.
 
 Going from data on spectra to more fundamental level, one now has 
 the shape and magnitude of an effective potential,
 as well as perturbative corrections to it, including spin-spin forces. Those can be
 used in wider context, in studies of QGP at high $T$.
 
  We do not provide quantitative results for the strip $T_c>T>1\, GeV$, but make a suggestion
  that rapid changes of the screening masses in it should be related to rapid changes of the
  density of the topological $L$-instanton dyons \cite{DeMartini:2021xkg}, the one which possesses fermionic zero modes and causes quark-antiquark attraction of the NJL type.

%%%%%%%%%%%%%%%%%%%%
%\bibliography{screening}

\begin{thebibliography}{99}

%\cite{DeTar:1987ar}
\bibitem{DeTar:1987ar}
C.~E.~Detar and J.~B.~Kogut,
%``The Hadronic Spectrum of the Quark Plasma,''
Phys. Rev. Lett. \textbf{59}, 399 (1987)
doi:10.1103/PhysRevLett.59.399
%203 citations counted in INSPIRE as of 07 Dec 2022

%\cite{Koch:1992nx}
\bibitem{Koch:1992nx}
V.~Koch, E.~V.~Shuryak, G.~E.~Brown and A.~D.~Jackson,
%``The Propagation of quarks in the spatial direction in hot QCD,''
Phys. Rev. D \textbf{46}, 3169 (1992)
[erratum: Phys. Rev. D \textbf{47}, 2157 (1993)]
doi:10.1103/PhysRevD.46.3169
[arXiv:hep-ph/9204236 [hep-ph]].
%74 citations counted in INSPIRE as of 06 Dec 2022

%\cite{Glozman:2019fku}
\bibitem{Glozman:2019fku}
L.~Y.~Glozman,
%``Three regimes of QCD,''
Int. J. Mod. Phys. A \textbf{36}, no.25, 2044031 (2021)
doi:10.1142/S0217751X20440315
[arXiv:1907.01820 [hep-ph]].
%8 citations counted in INSPIRE as of 29 Nov 2022

%\cite{Glozman:2022lda}
\bibitem{Glozman:2022lda}
L.~Y.~Glozman, O.~Philipsen and R.~D.~Pisarski,
%``Chiral spin symmetry and the QCD phase diagram,''
Eur. Phys. J. A \textbf{58}, no.12, 247 (2022)
doi:10.1140/epja/s10050-022-00895-4
[arXiv:2204.05083 [hep-ph]].
%9 citations counted in INSPIRE as of 19 Dec 2022

%\cite{Bazavov:2019www}
\bibitem{Bazavov:2019www} 
  A.~Bazavov {\it et al.},
  %``Meson Screening Masses in (2+1)-Flavor QCD,''
  arXiv:1908.09552 [hep-lat].
  %%CITATION = ARXIV:1908.09552;%%


%\cite{Laudicina:2022ghk}
\bibitem{Laudicina:2022ghk}
D.~Laudicina, M.~Dalla Brida, L.~Giusti, T.~Harris and M.~Pepe,
%``QCD mesonic screening masses and restoration of chiral symmetry at high T,''
[arXiv:2212.02167 [hep-lat]].
%0 citations counted in INSPIRE as of 20 Dec 2022

%\cite{Shuryak:1977ut}
\bibitem{Shuryak:1977ut}
E.~V.~Shuryak,
%``Theory of Hadronic Plasma,''
Sov. Phys. JETP \textbf{47}, 212-219 (1978)
IYF-77-34.
%371 citations counted in INSPIRE as of 20 Dec 2022

 
%\cite{Laine:2003bd}
\bibitem{Laine:2003bd}
M.~Laine and M.~Vepsalainen,
%``Mesonic correlation lengths in high temperature QCD,''
JHEP \textbf{02}, 004 (2004)
doi:10.1088/1126-6708/2004/02/004
[arXiv:hep-ph/0311268 [hep-ph]].
%72 citations counted in INSPIRE as of 06 Dec 2022

%\cite{DeMartini:2021xkg}
\bibitem{DeMartini:2021xkg}
D.~DeMartini and E.~Shuryak,
%``Chiral symmetry breaking and confinement from an interacting ensemble of instanton dyons in two-flavor massless QCD,''
Phys. Rev. D \textbf{104} (2021) no.9, 094031
doi:10.1103/PhysRevD.104.094031
[arXiv:2108.06353 [hep-ph]].
%5 citations counted in INSPIRE as of 21 Dec 2022
\end{thebibliography}
\end{document}